\documentclass[aip,jcp,reprint,twocolumn,amsmath,amssymb,floatfix,citeautoscript,nofootinbib]{revtex4-1}
\usepackage{cancel}
\usepackage{amsmath}
\usepackage{physics}
\usepackage{graphicx}
\usepackage{amssymb}
\usepackage{amsthm}
\usepackage{bm}
\usepackage{dcolumn}% Align table columns on decimal point
\newcolumntype{x}[1]{D{.}{.}{#1}}

\usepackage{ragged2e}
\usepackage{txfonts}
\allowdisplaybreaks

\usepackage{color}
\definecolor{myblue}{rgb}{0,0,1}
\usepackage[breaklinks=true,colorlinks=true,linkcolor=myblue,urlcolor=myblue,citecolor=myblue]{hyperref}

\usepackage{mathtools}
\usepackage{braket}% Enables braket notation
\usepackage{longtable}
\usepackage{multirow}
\usepackage[version=3]{mhchem}
\usepackage[capitalise]{cleveref}
\crefname{equation}{eqn}{eqns}
\Crefname{equation}{Eqn}{Eqns}
\crefname{figure}{Fig.}{Figs.}
\Crefname{figure}{Fig.}{Figs.}
\crefname{table}{Table}{Tables}
\Crefname{table}{Table}{Tables}
\definecolor{red}{RGB}{153,0,0}

\newcommand{\wavenumber}{\textrm{cm}^{-1}}

\begin{document}

% \title{
% Comment on ``A coming of age for many-body methods: Achieving consensus with experiments for CO on MgO''
% }

\title{
Adsorption and Vibrational Spectroscopy of CO on the Surface of MgO from Periodic Local Coupled-Cluster Theory
}

% Longer version
% \title{
% CO Adsorption on MgO Surface from Periodic Local Natural Orbital-Coupled Cluster with Singles, Doubles, and Perturbative Triples [LNO-CCSD(T)]
% }

\author{Hong-Zhou Ye}
\email{hzyechem@gmail.com}
\affiliation{Department of Chemistry, Columbia University, New York, NY 10027 USA}
\author{Timothy C. Berkelbach}
\email{t.berkelbach@columbia.edu}
\affiliation{Department of Chemistry, Columbia University, New York, NY 10027 USA}
\affiliation
{Initiative for Computational Catalysis, Flatiron Institute, New York, NY 10010, USA}

\begin{abstract}
    The adsorption of CO on the surface of MgO has long been a model problem in surface chemistry.
    Here, we report periodic Gaussian-based calculations for this problem using second-order perturbation theory (MP2) and coupled-cluster theory with single and double excitations (CCSD) and perturbative triple excitations [CCSD(T)], with the latter two performed using a recently developed extension of the local natural orbital approximation to problems with periodic boundary conditions.
    The low cost of periodic local correlation calculations allows us to calculate the full CCSD(T) binding curve of CO approaching the surface of MgO (and thus the adsorption energy) and the two-dimensional potential energy surface (PES) as a function of the distance from the surface and the CO stretching coordinate.
    From the PES, we obtain the fundamental vibrational frequency of CO on MgO, whose shift from the gas phase value is a common experimental probe of surface adsorption.
    We find that CCSD(T) correctly predicts a positive frequency shift upon adsorption of $+14.7~\wavenumber$, in excellent agreement with the experimental shift of $+14.3~\wavenumber$.
    We use our CCSD(T) results to assess the accuracy of MP2, CCSD, and several density functional theory (DFT) approximations, including exchange correlation functionals and dispersion corrections.
    We find that MP2 and CCSD yield reasonable binding energies and frequency shifts, whereas many DFT calculations overestimate the magnitude of the adsorption energy by $5$ -- $15$~kJ/mol and predict a negative frequency shift of about $-20~\wavenumber$, which we attribute to self-interaction-induced delocalization errors that are mildly ameliorated with hybrid functionals.
    Our findings highlight the accuracy and computational efficiency of the periodic local correlation for the simulation of surface chemistry with accurate wavefunction methods.
\end{abstract}

\maketitle

\section{Introduction}

The adsorption of a molecule on to the surface of a substrate is the first step of many interfacial physicochemical processes such as heterogeneous catalysis~\cite{Kroes21PCCP,Ye21ACS} and gas storage and separation~\cite{Sumida12CR,Qazvini21NCom,Chang22CM}.
Accurate determination of the adsorption energy, and more generally the potential energy surface (PES), is thus vital for understanding the structure and dynamics of the adsorption process, which in turn facilitates the interpretation of relevant interfacial experiments (e.g.,~molecular beam~\cite{Kleyn03CSR,Wang17PNAS,Kroes21PCCP},
temperature-programmed desorption~\cite{Falconer83CR,Wichtendahl99PSSA,Sterrer06ACA,Dohnalek01JPCB},
and surface vibrational spectroscopies~\cite{Zhang08JACS,Geiger09ARPC}).
Nonetheless, this has been proven a challenging computational task over the past two decades~\cite{Schimka10NM,Sauer19ACR}.
The major difficulty stems from the inherently weak yet correlated nature of surface-adsorbate interactions, which necessitates achieving sub-kJ/mol accuracy for reliable comparison with experiments~\cite{Schimka10NM,Boese13PCCP,Sauer19ACR,Shi23JACS} and development of accurate force fields for molecular dynamics~\cite{Zhang18PRL,Unke21CR,Wen23PNAS,Zeng23NC}.
The demanded high accuracy is typically out of reach by density functional theory~\cite{Hohenberg64PR} (DFT) whose performance can depend sensitively on the choice of approximate exchange-correlation functionals (especially because of self-interaction errors) and dispersion correction methods~\cite{Hensley17JPCC,Rehak20PCCP,Smeets21JPCC,Meng21JPCC,Kroes21PCCP}.

In principle, correlated wavefunction theories that include the correct many-body physics hold promise for computational surface science.
In practice, however, their applicability is often hindered by the high computational cost required to reach convergence with respect to the one-particle basis set size, the supercell and slab size, and the level of electron correlation treatment.
In this work, we focus on CO adsorption on the MgO (001) surface, which is an extensively studied model system for high levels of theory~\cite{Boese13PCCP,Maristella19JCTC,Mitra22JPCL,Shi23JACS}.
Recently, Shi and co-workers~\cite{Shi23JACS} applied coupled-cluster theory with singles, doubles, and perturbative triples~\cite{Raghavachari89CPL} [CCSD(T)], commonly known as the ``gold standard'' of main-group quantum chemistry, within an embedded-cluster framework~\cite{Shi22JCP} to produce their best estimate of the adsorption energy, $-19.2 \pm 1.0$~kJ/mol, which was in good agreement with that calculated by canonical periodic CCSD(T) and diffusion Monte Carlo, as well as the value estimated from temperature-programmed desorption experiments~\cite{Dohnalek01JPCB} (about $-20.0$~kJ/mol).

In a preprint version of the present manuscript~\cite{Ye23MgOpreprint}, we applied our recently developed periodic local CCSD(T) method~\cite{Ye23underreview,Ye23inprep} to calculate an adsorption energy in good agreement with the above values.
Due to the use of local correlation theory and recent developments in periodic integral evaluation~\cite{Ye21JCPa,Ye21JCPb,Bintrim22JCTC} and correlation-consistent Gaussian basis sets~\cite{Ye22JCTC}, these calculations were relatively affordable: they were generated from scratch in a few days, requiring about $18$k CPU hours.
% Because of this low cost, we now extend our work to calculate a smooth, two-dimensional PES for CO on MgO with CCSD(T).
Because of this low cost, we now extend our work to calculate a smooth, two-dimensional PES for CO on MgO, from which we calculate the vibrational frequency shift of CO stretching upon adsorption to be $+14.7~\wavenumber$, which is in nearly quantitative agreement with the infrared experimental result~\cite{Spoto03SS} of $+14.3~\wavenumber$.
Using the CCSD(T) PES, we assess the accuracy of lower-level wavefunction theories and various DFT protocols that combine commonly used exchange-correlation functionals and dispersion corrections with respect to the calculated binding curve (\cref{subsec:binding_curve}) and vibrational frequency shift of CO stretching upon adsorption (\cref{subsec:vibrational_shift}).
% Notably, none of the tested DFT protocols achieves both accurate energy and accurate vibrational frequency shift.
% By contrast, second-order M{\o}ller-Plesset perturbation theory (MP2) reproduces the entire CCSD(T) PES within approximately $1$ kJ/mol of accuracy \red{and predicts a vibrational frequency shift in agreement with CCSD(T) within about $1~\wavenumber$}.
% We then evaluate the vibrational frequency shift of CO stretching upon adsorption using CCSD(T) and find good agreement with infrared experiments.
We find that none of the tested DFT protocols can achieve high accuracy in both the binding energy and the vibrational frequency.
By contrast, MP2 reproduces the entire CCSD(T) PES within about 1~kJ/mol and the vibrational frequency shift within about $1~\wavenumber$, while CCSD performs slightly worse with an error of about $4$~kJ/mol and $3~\wavenumber$, respectively.
We conclude the work in \cref{sec:conclusion} with a discussion on potential future developments.

\begin{figure*}[!t]
    \centering
    \includegraphics[width=0.98\linewidth]{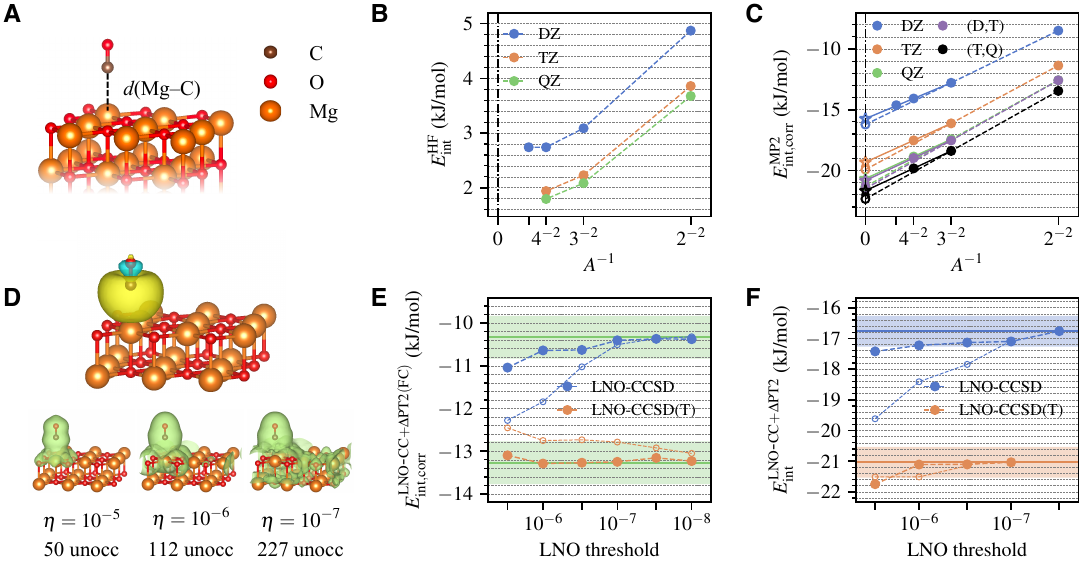}
    \caption{(A) Atomic structures of a CO molecule physically adsorbed on a MgO (001) surface.
    The Mg-C distance is highlighted.
    (B) Convergence of the HF interaction energy with surface and basis set sizes.
    (C) Convergence of the correlation part of the MP2 interaction energy with surface and basis set sizes.
    The TDL results estimated by extrapolating the surface size with $A = 2\times2$ and $3\times3$ and with $A = 3\times3$ and $4\times4$ are shown as dashed and solid lines, respectively.
    The CBS results from (DZ,TZ) and (TZ,QZ) extrapolation are also shown.
    (D) Isosurface plot of a representative localized occupied orbital on the CO molecule (obtained using the generalized Pipek-Mezey method~\cite{Pipek89JCP,Lehtola14JCTC}) and the density of the corresponding unoccupied LNOs with tightening truncation threshold $\eta$ (generated for a $3\times3$ surface using the TZ basis set).
    The number of unoccupied LNOs are listed at the bottom, which is only a fraction of the total number of unoccupied orbitals, $1083$.
    (E) Convergence of the correlation part of the frozen-core LNO-CCSD/CCSD(T) interaction energy with the LNO truncation threshold $\eta$ for a $2\times2$ surface with (DZ,TZ)-extrapolated CBS limit.
    Both the regular LNO-CCSD/CCSD(T) results (hollow circles) and those corrected using \cref{eq:lnocc_cano_correction} (filled circles) are shown.
    The corrected LNO-CCSD/CCSD(T) interaction energies are seen to converge much faster to the canonical CCSD/CCSD(T) results, as indicated by the green horizontal line (the shaded area indicates an error of $\pm 0.5$~kJ/mol).
    (F) Convergence of the correlation part of the final LNO-CCSD/CCSD(T) interaction energy with the LNO truncation threshold $\eta$.
    The (DZ,TZ) extrapolated CBS limit is employed.
    Results from two different schemes for the TDL extrapolation, $A = 2\times2$ and $3\times3$ (filled circles) and $A = 3\times3$ and $4\times4$ (hollow circles), agree well at convergence.
    The converged LNO-CCSD/CCSD(T) values are indicated by the solid horizontal line, with the shaded area indicating an error of $\pm 0.5$~kJ/mol.
    }
    \label{fig:conv}
\end{figure*}

\section{Computational Details}

The system being studied is shown in \cref{fig:conv}A.
A CO molecule is adsorbed with its C end on a five-fold coordinated Mg site of the pristine MgO (001) surface.
The physisorption results in only slight perturbations to the surface and negligible change of the CO bond length.
We model the surface using a two-layer MgO slab, which is sufficient to converge the calculated adsorption energy to accuracy within $0.3$~kJ/mol (Table S5).
The equilibrium geometries for CO, MgO, and MgO+CO are obtained using DFT with the Perdew-Burke-Ernzerhof (PBE) functional~\cite{Perdew96PRL} and the D3 dispersion correction~\cite{Grimme10JCP} using Quantum Espresso~\cite{Giannozzi09JPCM,Giannozzi17JPCM}.
Atoms in the bottom layer of the MgO slab are fixed during geometry optimization.

We calculate the adsorption energy of CO on MgO as follows
\begin{equation}    \label{eq:Eads}
    E_{\textrm{ads}}
        = E_{\textrm{int}} + \Delta_{\textrm{geom}}
\end{equation}
where $E_{\textrm{int}}$ is the interaction energy between CO and MgO calculated with their respective geometries fixed to be those in the MgO+CO composite system,
and $\Delta_{\textrm{geom}}$ is the geometry relaxation energy.
Following ref~\citenum{Shi23JACS}, we use our DFT-determined $\Delta_{\textrm{geom}}$ ($1.1$~kJ/mol from PBE+D3, which is in good agreement with previous work~\cite{Boese13PCCP,Maristella19JCTC,Shi23JACS}) in \cref{eq:Eads} and evaluate $E_{\textrm{int}}$ using correlated wavefunction methods.

All wavefunction calculations are performed with the PySCF code~\cite{Sun18WIRCMS,Sun20JCP}.
The GTH-cc-pV$X$Z basis sets~\cite{Ye22JCTC}, augmented by diffuse functions for the CO molecule and nearby surface atoms, are employed with the GTH pseudopotential optimized for HF~\cite{Goedecker96PRB,Hartwigsen98PRB,HutterPP}.
All interaction energies are corrected for basis set superposition error.
The two-electron integrals are treated by the range-separated density fitting algorithm~\cite{Ye21JCPa,Ye21JCPb}, and the integrable divergence of the HF exchange is treated using a Madelung constant correction~\cite{Paier06JCP,Broqvist09PRB,Sundararaman13PRB}.
The $O(N^6)$/$O(N^7)$ cost scaling of canonical CCSD/CCSD(T) is avoided by a periodic extension of the local natural orbital (LNO) approximation, which we have recently applied to study the dissociation of water on the surface of \ce{Al2O3} and \ce{TiO2}~\cite{Ye23underreview}.
Our periodic implementation, detailed in ref~\citenum{Ye23underreview}, closely follows the molecular LNO-CCSD/CCSD(T) originally developed by K{\'a}llay and co-workers~\cite{Rolik11JCP,Rolik13JCP}.
We calculate the correlation energy of a supercell containing $N$ electrons as a sum of contributions from all $N$ localized occupied orbitals $i$,
\begin{equation}
    E^{\textrm{LNO-CC}}_{\textrm{corr}}(\eta)
        = \sum_{i}^{N} E^{\textrm{LNO-CC}}_{i,\textrm{corr}}(\eta).
\end{equation}
Each local contribution $E^{\textrm{LNO-CC}}_{i,\textrm{corr}}$ is evaluated independently in a truncated set of occupied and unoccupied LNOs that are optimized for local orbital $i$~\cite{Rolik11JCP,Rolik13JCP}.
The accuracy of the LNO approximation can be systematically improved by adjusting a single parameter: the threshold $\eta$ used to truncate the LNOs by their occupation numbers (following ref~\citenum{Rolik13JCP}, we use a threshold for unoccupied orbitals that is fixed to be $10$ times smaller than that of the occupied orbitals; all reported thresholds henceforth are for unoccupied orbitals).
A representative localized occupied orbital is shown in \cref{fig:conv}D, together with the density of the corresponding unoccupied LNOs, whose spatial extension increases as the truncation threshold tightens.
As $\eta \to 0$, the LNO-CCSD/CCSD(T) results converge to the result of canonical CCSD/CCSD(T) calculations.
To expedite this convergence, we apply a standard MP2 correction evaluated at the same $\eta$~\cite{Rolik11JCP}.
This is equivalent to correcting full MP2 by a short-range CC correction evaluated within the LNO subspace, i.e.,
% \begin{equation}    \label{eq:lno_pt2_corr}
%     E^{\textrm{LNO-CC+}\Delta\textrm{PT2,corr}}(\eta)
%         = E^{\textrm{MP2,corr}} + \Delta E^{\textrm{LNO-CC,corr}}(\eta)
% \end{equation}
\begin{equation}    \label{eq:lno_pt2_corr}
    E^{\textrm{LNO-CC}+\Delta\textrm{PT2}}_{\textrm{corr}}(\eta)
        = E^{\textrm{MP2}}_{\textrm{corr}} + \Delta E^{\textrm{LNO-CC}}_{\textrm{corr}}(\eta)
\end{equation}
where
\begin{equation}    \label{eq:lnocc_correction}
    \Delta E^{\textrm{LNO-CC}}_{\textrm{corr}}(\eta)
        % = E^{\textrm{LNO-CC,corr}}(\eta) - E^{\textrm{LNO-MP2,corr}}(\eta)
        = \sum_{i}^{N} E^{\textrm{LNO-CC}}_{i,\textrm{corr}}(\eta)
        - E^{\textrm{LNO-MP2}}_{i,\textrm{corr}}(\eta)
\end{equation}
\Cref{eq:lno_pt2_corr} is also used to account for other effects such as core-electron correlation.
More computational details can be found in the Supplementary Information.

\section{CO on MgO (001) surface}

\subsection{Equilibrium adsorption energy}
\label{subsec:eq_ads_energy}

We first validate our periodic LNO-CC methodology by calculating the equilibrium adsorption energy $E_{\textrm{ads}}$ for a single CO molecule adsorbed on the MgO (001) surface.
Reference values of $E_{\textrm{ads}}$ can be derived from the desorption activation energy measured by temperature-programmed desorption experiments~\cite{Wichtendahl99PSSA,Sterrer06ACA,Dohnalek01JPCB}.
This was done earlier by Boese and Sauer~\cite{Boese13PCCP} who obtained a reference value of $-20.6 \pm 2.4$~kJ/mol based on the experimental work by Dohn\'{a}lek \textit{et al.}~\cite{Dohnalek01JPCB}, and more recently by Shi and co-workers~\cite{Shi23JACS} who derived a modified value of $-19.2 \pm 1.0$~kJ/mol by averaging two experimental results.~\cite{Wichtendahl99PSSA,Dohnalek01JPCB}
A correct theoretical prediction of $E_{\textrm{ads}}$, however, necessitates the use of correlated wavefunction methods.
Boese and Sauer~\cite{Boese13PCCP} employed cluster-based MP2-in-DFT embedding and found an $E_{\textrm{ads}}$ of $-20.9 \pm 0.7$~kJ/mol, and
Shi and co-workers~\cite{Shi23JACS} calculated $E_{\textrm{ads}}$ using embedded-cluster LNO-CCSD(T) to be $-19.2 \pm 1.0$~kJ/mol.
Clearly, both values are in good agreement with experimental estimates.

We revisit this problem to establish a protocol to achieve sub-kJ/mol accuracy for the CCSD(T) interaction energy.
We use a fixed $d(\textrm{Mg-C}) = 2.460$~\AA{} for this purpose, which also facilitates comparison with previous works~\cite{Boese13PCCP,Mazheika16JPCC,Mitra22JPCL,Shi23JACS}.
Using \cref{eq:lno_pt2_corr}, the CC interaction energy can be calculated as
\begin{equation}    \label{eq:Eint_lnoccpt2}
    % E_{\textrm{int}}^{\textrm{LNO-CC+}\Delta\textrm{PT2}}(A,X,\eta)
    E_{\textrm{int}}^{\textrm{CC}}
        = \lim_{\substack{A \to \infty \\ X \to \infty \\ \eta \to 0}}
        E_{\textrm{int}}^{\textrm{MP2}}(A,X)
        + \Delta E_{\textrm{int,corr}}^{\textrm{LNO-CC(FC)}}(A,X,\eta)
\end{equation}
where $E_{\textrm{int}}^{\textrm{MP2}}$ is the MP2 interaction energy and $\Delta E_{\textrm{int,corr}}^{\textrm{LNO-CC(FC)}}$ is the LNO-CC correction (\ref{eq:lnocc_correction});
we evaluate the latter with the $[2s^2 2p^6]$ semicore electrons of Mg being frozen.
Both terms in \cref{eq:Eint_lnoccpt2} need to be converged to the thermodynamic limit (TDL) with respect to the surface size $A$ and to the complete basis set (CBS) limit with the basis set size $X$, while $\Delta E_{\textrm{int,corr}}^{\textrm{LNO-CC(FC)}}$ also needs to be converged to the full CC limit with the LNO truncation threshold $\eta$.

% \section{Results and discussion}
%
% We first establish a protocol to achieve sub-kJ/mol accuracy for the CCSD(T) interaction energy.
% We use a fixed $d = 2.460$~\AA{} for this purpose, which also facilitates comparison with previous works~\cite{Boese13PCCP,Mazheika16JPCC,Mitra22JPCL,Shi23JACS}.
% $E_{\textrm{int}}(A,L,X)$ needs to be converged with respect to the surface size $A$, the slab thickness $L$, and the basis set size $X$.
% As shown in TABLE S5, slabs with two atomic layers are sufficient for converging $E_{\textrm{int}}$ with respect to $L$ to about $3$~meV.
% In what follows, we thus fix $L=2$ and focus on converging $E_{\textrm{int}}(A,X)$ to the TDL and the CBS limit, denoted as $A_{\infty}$ and $X_{\infty}$, respectively.

The MP2 interaction energy can be further decomposed as a sum of the HF part ($E_{\textrm{int}}^{\textrm{HF}}$) and the MP2 correlation part ($E_{\textrm{int,corr}}^{\textrm{MP2}}$).
\Cref{fig:conv}B shows the convergence of $E_{\textrm{int}}^{\textrm{HF}}(A,X)$.
We see that using a $4\times4$ surface and a QZ basis set essentially attains both the TDL and the CBS limit, giving $E_{\textrm{int}}^{\textrm{HF}}(A_{\infty},X_{\infty}) \approx 1.8 \pm 0.1$~kJ/mol, where half the difference between TZ and QZ results is taken as an estimate of the remaining basis set incompleteness error.
\Cref{fig:conv}C shows the convergence of $E_{\textrm{int,corr}}^{\textrm{MP2}}(A,X)$, which is slower than that of $E_{\textrm{int}}^{\textrm{HF}}(A,X)$ in both parameters.
However, reliable extrapolations to both limits can be performed based on the asymptotic behaviors
\begin{subequations}    \label{eq:MP2_corr_extrapolation}
\begin{align}
    E_{\textrm{int,corr}}^{\textrm{MP2}}(A,X)
        &\approx E_{\textrm{int,corr}}^{\textrm{MP2}}(A_{\infty},X) + c_1 A^{-1}
    \label{subeq:mp2_extrap_A}
    \\
    E_{\textrm{int,corr}}^{\textrm{MP2}}(A,X)
        &\approx E_{\textrm{int,corr}}^{\textrm{MP2}}(A,X_{\infty}) + c_2 X^{-3}
    \label{subeq:mp2_extrap_X}
\end{align}
\end{subequations}
A good estimate of $E_{\textrm{int,corr}}^{\textrm{MP2}}(A_{\infty},X_{\infty})$ is obtained by extrapolating the surface size with $A=3\times3$ and $4\times4$ and the basis set size with TZ ($X=3$) and QZ ($X=4$).
This gives \(
    E_{\textrm{int,corr}}^{\textrm{MP2}}(A_{\infty},X_{\infty})
        \approx -21.7 \pm 0.3
\)~kJ/mol, where the uncertainty arises from the remaining basis set incompleteness error and the pseudopotential error (TABLE S2).
Notably, repeating the calculations without correlating Mg's semicore electrons gives $E_{\textrm{int,corr}}^{\textrm{MP2(FC)}}(A_{\infty},X_{\infty}) \approx -19.5$~kJ/mol, revealing a sizable contribution of about $-2.2$~kJ/mol from the semicore electrons.
Summing the HF and MP2 correlation contributions, we arrive at our final estimate of $E_{\textrm{int}}^{\textrm{MP2}}$, which is $-19.9 \pm 0.3$~kJ/mol.

Transitioning to CC, we first show in \cref{fig:conv}E the $\eta$-convergence of $E_{\textrm{int,corr}}^{\textrm{LNO-CC+}\Delta\textrm{PT2(FC)}}(A,X,\eta)$, extrapolated to the CBS limit based on DZ and TZ results for a $2 \times 2$ surface.
The system is small enough (about $650$ orbitals in the TZ basis set) to allow performing canonical CC calculations, whose results are also shown in \cref{fig:conv}E for comparison.
We see that the error of LNO-CCSD(T) (orange hollow circles) drops fast to $0.5$~kJ/mol with a modest threshold of $10^{-6}$, but achieving a similar accuracy in LNO-CCSD (blue hollow circles) requires a tighter threshold of $10^{-7}$.
% We see that the error of LNO-CCSD and LNO-CCSD(T) (hollow circles) is about $1.5$ and $0.5$~kJ/mol with a modest threshold of $10^{-6}$, but an error of $0.5$~kJ/mol requires a tighter threshold of $10^{-7}$.
While LNO-CC calculations with a tight threshold of $10^{-7}$ are feasible for larger systems, a more efficient alternative is to correct the LNO-CC results by the difference from an inexpensive canonical CC calculation using DZ basis set and $2 \times 2$ surface, which we denoted as
\begin{equation}    \label{eq:lnocc_cano_correction}
\begin{split}
    \Delta^{\textrm{CC(FC)}}(\eta)
        = E_{\textrm{int,corr}}^{\textrm{CC(FC)}}(2\times2, \textrm{DZ})
        - E_{\textrm{int,corr}}^{\textrm{LNO-CC(FC)}}(2\times2, \textrm{DZ}, \eta)
\end{split}
\end{equation}
The corrected LNO-CC results are shown as filled circles in \cref{fig:conv}E and seen to converge much faster than the uncorrected ones.
In FIG.~S2, we confirmed that the same correction (\ref{eq:lnocc_cano_correction}) is also transferrable for correcting larger surfaces.

To obtain our final estimate of the CC interaction energy using \cref{eq:Eint_lnoccpt2}, we extrapolate $\Delta E_{\textrm{int,corr}}^{\textrm{LNO-CC(FC)}}(A,X,\eta)$ to the TDL and the CBS limit using $A = 2\times2$ and $3\times3$ and the DZ and TZ basis sets, and add this number to the converged $E_{\textrm{int}}^{\textrm{MP2}}$ determined above.
The resulting $\eta$-dependent LNO-CC interaction energy is then converged with respect to $\eta$ (\cref{fig:conv}F).
Our best estimate of the CCSD and CCSD(T) interaction energy, obtained with $\eta = 10^{-7}$, is $-16.8 \pm 0.5$~kJ/mol and $-21.0 \pm 0.5$~kJ/mol,
where the error bar combines that of $E_{\textrm{int}}^{\textrm{MP2}}$ ($0.3$~kJ/mol), of $\Delta^{\textrm{CC(FC)}}$ ($0.2$~kJ/mol), and of the LNO truncation error which is estimated to be $0.3$~kJ/mol using the difference between the two LNO-CCSD calculations with the tightest thresholds shown in \cref{fig:conv}F.

Our final numbers for the HF, MP2, CCSD, and CCSD(T) $E_{\textrm{ads}}$ are summarized in \cref{tab:final}, where they are compared to both the experimental reference and other computational results from literature.
Comparing to ref~\citenum{Shi23JACS}, our periodic Gaussian-based approach shows sub-kJ/mol agreement with the embedded-cluster approach for both the MP2 and CCSD(T) adsorption energy, and a slightly worse agreement of $2$~kJ/mol with the periodic plane-wave CCSD and CCSD(T), likely due to the larger error bar associated with the latter.
Given the use of slightly different geometries and different strategies to attain convergence along the necessary theoretical axes, the level of agreement achieved between our work and ref~\citenum{Shi23JACS} is remarkable.
Our final CCSD(T) adsorption energy, $-20.0 \pm 0.5$~kJ/mol, also agrees well with corrected~\cite{Shi23JACS,Boese13PCCP} temperature-programmed desorption experimental results~\cite{Wichtendahl99PSSA,Dohnalek01JPCB}.
By contrast, insufficient convergence along one or more of the theoretical axes can lead to significant errors in the calculated interaction energy~\cite{Mazheika16JPCC,Mitra22JPCL}, which has been discussed thoroughly in previous work~\cite{Shi23JACS}. 

%%%%% Single-column table with Eads (normal column), Ref
%%%%%
\begin{table}[t]
    \centering
    \caption{CO adsorption energy $E_{\textrm{ads}}$ on the MgO (001) surface with a Mg-C distance of $2.460$~\AA{}~obtained through various wavefunction methods compared to recent results in literature.
    The error bars of our numbers are explained in the main text, while those of the results in ref~\citenum{Shi23JACS} have the uncertainty in $\Delta_{\textrm{relax}}$ removed.
    }
    \label{tab:final}
    \begin{ruledtabular}
        \def\arraystretch{1.1}%  1 is the default, change whatever you need
    \begin{tabular}{llrl}
        Method & Comput. details\footnotemark[1] &
            $E_{\textrm{ads}}$~(kJ/mol) & Reference   \\
        \hline
        HF & periodic/GTO & $+\phantom{1}2.9 \pm \phantom{1}0.1$ & \multirow{4}*{this work}    \\
        MP2 & periodic/GTO & $-18.8 \pm \phantom{1}0.3$ &     \\
        CCSD & periodic/GTO/LNO & $-15.7 \pm \phantom{1}0.5$ &      \\
        CCSD(T) & periodic/GTO/LNO & $-20.0 \pm \phantom{1}0.5$ &      \\
        \hline
        MP2 & cluster/GTO & $-18.5 \pm \phantom{1}0.5$ & \multirow{5}*{ref~\citenum{Shi23JACS}}     \\
        CCSD(T) & cluster/GTO/LNO & $-19.2 \pm \phantom{1}0.6$ &     \\
        CCSD & periodic/PW/FNO & $-14.0 \pm \phantom{1}2.1$ &     \\
        CCSD(T) & periodic/PW/FNO & $-18.6 \pm \phantom{1}2.1$ &    \\
        DMC & periodic & $-18.1 \pm \phantom{1}2.3$ &    \\
        \hline
        \multirow{2}*{Experiments}
            & & $-20.6 \pm 2.4$\footnotemark[2] & ref~\citenum{Boese13PCCP}  \\
            & & $-19.2 \pm 1.0$\footnotemark[3] & ref~\citenum{Shi23JACS}  \\
    \end{tabular}
    \end{ruledtabular}
    \footnotetext[1]{Acronyms: Gaussian-type orbital (GTO), plane wave (PW), frozen natural orbital (FNO), diffusion Monte Carlo (DMC), density matrix embedding theory (DMET).}
    \footnotetext[2]{Based on TPD experiments in ref~\citenum{Dohnalek01JPCB}.}
    \footnotetext[3]{Based on TPD experiments in ref~\citenum{Dohnalek01JPCB,Wichtendahl99PSSA}.}
\end{table}

\subsection{Full binding curve}
\label{subsec:binding_curve}

\begin{figure*}[!t]
    \centering
    \includegraphics[width=0.8\linewidth]{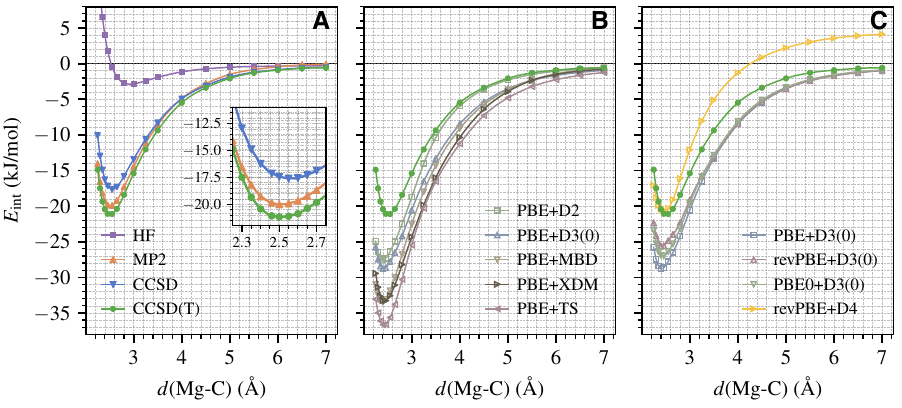}
    \caption{The full binding curve for CO adsorption on the MgO (001) surface calculated using selected (A) wavefunction and (B,C) DFT methods.
    The energy data are denoted by the markers.
    The solid curves are obtained by fitting the data to the Morse potential.
    The insert in panel (A) shows a zoom-in view for $d(\textrm{Mg-C})$ ranging from $2.25$ to $2.75$~\AA{}.}
    \label{fig:pes_diss}
\end{figure*}

We now employ the established protocol to calculate the full binding curve for CO adsorption on the MgO (001) surface.
An accurate binding curve is important for understanding the adsorption/desorption dynamics~\cite{Smeets19JPCA}, which can be directly related to molecular beam experiments~\cite{Kroes21PCCP}.
Here, we focus on calculating the variation of $E_{\textrm{int}}$ as a function of $d(\textrm{Mg-C})$ (with atoms in the surface kept frozen).
The results from the wavefunction methods discussed in \cref{subsec:eq_ads_energy} are shown in \cref{fig:pes_diss}A.
We see that our protocol generates smooth energy curves for all methods including CCSD(T).
This is remarkable given the small energy scale involved here (about~$2$~kJ/mol near equilibrium; see the insert of \cref{fig:pes_diss}A) and highlights the extremely high fidelity of our periodic LNO-CCSD(T) methodology.
Performance-wise, CCSD and MP2 underestimate the CCSD(T) binding energy by about $4$ and $1$~kJ/mol for a wide range of $d(\textrm{Mg-C})$ near the equilibrium geometry, which is consistent with the trend discussed in \cref{subsec:eq_ads_energy} for the equilibrium $E_{\textrm{ads}}$.
All three methods smoothly approach the correct dissociation limit.
Interestingly, the HF binding curve does show a shallow well of about $-3$~kJ/mol at a large $d(\textrm{Mg-C})$ of nearly $3$~\AA{}.

Armed with CCSD(T) results, we can now assess the performance of different DFT methods.
\Cref{fig:pes_diss}B presents the binding curves obtained using the PBE functional with five popular dispersion correction models: the exchange-hole dipole moment (XDM) model~\cite{Becke05JCP}, the Tkatchenko-Scheffler (TS) model~\cite{Tkatchenko09PRL}, the many-body dispersion (MBD) model~\cite{Tkatchenko12PRL},
and two different generations of Grimme's DFT-D model (D2~\cite{Grimme06JCC} and D3~\cite{Grimme10JCP}).
We see that all five DFT protocols significantly overestimate the equilibrium binding energy by about $6$~kJ/mol (D2) to $16$~kJ/mol (TS) and predict the equilibrium Mg-C separation to be too small by 5--10~pm.
To investigate the effect of different exchange-correlation functionals, in \cref{fig:pes_diss}C we present binding curves calculated using PBE and its two extensions, revPBE~\cite{Zhang98PRL} and PBE0~\cite{Adamo99JCP}, all corrected by the D3 dispersion correction.
It is evident that both revPBE and PBE0 ameliorate PBE's overbinding behavior, but the improvement is way too small.
As seen in \cref{fig:pes_diss}C, we find that revPBE+D4~\cite{Caldeweyher17JCP} gives the best performance
near the equilibrium geometry, in agreement with ref~\citenum{Shi23JACS}, but the D4 dispersion energy does not decay properly to zero in the dissociation limit.

\subsection{Frequency shift of CO stretching}
\label{subsec:vibrational_shift}

\begin{figure*}
    \centering
    \includegraphics[width=0.98\linewidth]{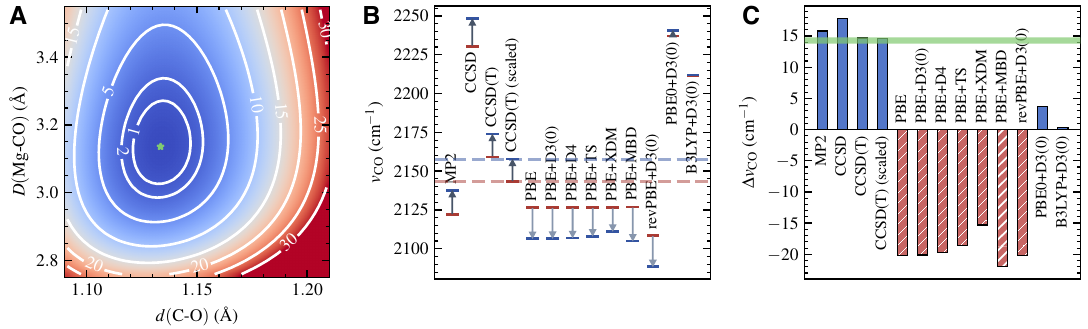}
    \caption{(A) CCSD(T) potential energy surface for the CO stretching mode coupled with its adsorption onto the MgO (001) surface, calculated using LNO-CCSD(T) with the protocol established in \cref{subsec:eq_ads_energy}.
    $d(\textrm{C-O})$ is the CO bond length and $D(\textrm{Mg-CO})$ is the distance between the CO center of mass and the surface Mg atom.
    The number labelling each contour line is the energy relative to the energy minimum (denoted by a green star) in kJ/mol.
    (B) The CO stretching frequency for a free CO molecule (red) and a CO molecule adsorbed on the MgO surface (blue) predicted by different methods.
    The scaled CCSD(T) frequencies, using a scaling factor based on our gas-phase calculation and experiment, as discussed in the main text, are also shown.
    The experimental values are shown as dashed horizontal lines of corresponding color.
    For each computational method, a gray arrow indicates the direction of the predicted frequency shift upon adsorption.
    (C) Shift of the CO stretching frequency upon adsorption calculated from the frequency data shown in panel (B).
    Solid blue and hatched red bars indicate blue and red shift, respectively.
    The experimental reference is shown as the green shaded area.}
    \label{fig:pes_vib}
\end{figure*}

We now employ the established protocol to calculate the vibrational frequency of CO on the MgO (001) surface.
Experimentally, this has been measured by infrared spectroscopy~\cite{Spoto03SS} to be $\nu_{\textrm{CO(ads)}} = 2157.5~\wavenumber$.
Compared to the gas-phase frequency of $\nu_{\textrm{CO(g)}} = 2143.2~\wavenumber$, this corresponds to a positive vibrational
frequency shift $\Delta \nu_{\textrm{CO}} = \nu_{\textrm{CO(ads)}} - \nu_{\textrm{CO(g)}} = 14.3~\wavenumber$.
Such frequency shifts are commonly used for probing both the structural and electronic information of a surface~\cite{Benziger82JC,Sterrer06ACA}.
Accurately resolving such a small frequency shift serves as a stringent test for electronic structure theory.

A minimal theoretical description necessitates calculating a two-dimensional potential energy surface for the CO bond length $d(\textrm{C-O})$ and the surface-CO distance, chosen here to be $D(\textrm{Mg-CO})$, the distance between the CO center of mass and the surface Mg atom (the surface is kept frozen at its equilibrium geometry following previous work~\cite{He23JPCA}).
We calculate this two-dimensional PES using various electronic structure methods by performing a series of single-point calculations over a total of $130$ different geometries and fitting the resulting energy data using a sixth-order polynomial.
The fitting error is found to be less than $0.5$~kJ/mol for all electronic structure methods discussed below.
We then extract $\nu_{\textrm{CO(ads)}}$ in the harmonic approximation from the second derivatives of the fitted PES.
We calculate $\nu_{\textrm{CO(g)}}$ in a similar manner from a one-dimensional PES of $d(\textrm{C-O})$ for a free CO molecule.
See Supplementary Information for more details.
% In particular, we scan $d(\textrm{C-O})$ from $0.982$ to $1.432$~\AA{} with an increment of $0.05$~\AA{} and $D(\textrm{Mg-CO})$ from $2.630$ to $3.830$~\AA{} with an increment of $0.1$~\AA{}, obtaining a total of $130$ data points.

The PES for the surface adsorbed CO obtained using LNO-CCSD(T) is shown in \cref{fig:pes_vib}A.
The contour of the PES displays a characteristic elliptical shape for the coupling of a strong vibrational mode to a weak one.
From the PES of the adsorbed CO and that of a free CO (not shown), we extract $\nu_{\textrm{CO(ads)}}^{\textrm{CCSD(T)}} = 2173.9~\wavenumber$ and  $\nu_{\textrm{CO(g)}}^{\textrm{CCSD(T)}} = 2159.2~\wavenumber$, both of which are about $16~\wavenumber$ higher than the experimentally observed frequencies.
As a result, the frequency shift predicted by CCSD(T), $\Delta \nu^{\textrm{CCSD(T)}}_{\textrm{CO}} = 14.7~\wavenumber$, is in quantitative agreement with experiment, with an error less than $0.5~\wavenumber$.
From a practical point of view, the harmonic CCSD(T) overestimation of vibrational frequencies of small molecules is well-known and usually corrected by applying an empirical scaling factor.~\cite{Scott96JPC,Tew07JPCA}
The nearly perfect agreement between the computational and experimental $\Delta \nu_{\textrm{CO}}$ thus indicates a highly transferrable scaling factor for CO in gas and condensed phases.
Indeed, applying the scaling factor $0.9926$ determined by our gas-phase calculation and the experiment, we obtain a scaled vibrational frequency for adsorbed CO, $\nu_{\textrm{CO(ads)}}^{\textrm{CCSD(T)-scaled}} = 2157.8~\wavenumber$, which agrees nearly perfectly with the experimental frequency.

The CCSD(T) frequency and frequency shift discussed above are displayed in \cref{fig:pes_vib}B and C.
In the same plots, we also show the absolute frequency and frequency shift predicted by several different electronic structure methods.
MP2 slightly underestimates $\nu_{\textrm{CO(ads)}}$ by about $20~\wavenumber$, while CCSD overestimates it by about $90~\wavenumber$.
However, both methods also predict a similar trend for $\nu_{\textrm{CO(g)}}$, making their frequency shift, $15.8~\wavenumber$ from MP2 and $17.8~\wavenumber$ from CCSD, in reasonable agreement with the experiment.

In contrast, DFT with semilocal functionals, such as PBE and revPBE, predicts a large negative $\Delta \nu_{\textrm{CO}}$ ranging from $-15$ to $-20~\wavenumber$, regardless of the dispersion correction being used (\cref{fig:pes_vib}C).
This qualitative error is caused by the imbalanced error in the adsorbed and gas-phase calculations: while both $\nu_{\textrm{CO(ads)}}$ and $\nu_{\textrm{CO(g)}}$ are underestimated, the former is underestimated to a much larger extent (\cref{fig:pes_vib}B), likely due to a larger delocalization error in condensed-phase systems.
Improvement is seen by using hybrid functionals such as PBE0 and B3LYP, which suffer less from the delocalization error due to partial inclusion of the HF exchange energy~\cite{Mardirossian17MP}.
However, the predicted $\Delta\nu_{\textrm{CO}}$, $3.7~\wavenumber$ from PBE0+D3(0) and $0.3~\wavenumber$ from B3LYP+D3(0), are still too small compared to the experiment.
Despite having qualitatively correct shapes, the PESs obtained using these DFT methods exhibit clear quantitative deviations from the CCSD(T) PES (FIG.~S5).
This highlights the challenge for achieving spectroscopic accuracy for vibrations of adsorbed molecules with commonly used DFT methods.

\section{Conclusion}
\label{sec:conclusion}

To summarize, we computationally studied the adsorption of a single CO molecule on the MgO (001) surface using the ``gold-standard'' method of quantum chemistry, CCSD(T), made possible by our recently developed periodic LNO-CCSD(T) method.
After validating our method using the equilibrium adsorption energy, we leverage the low computational cost and high fidelity of LNO-CCSD(T) to calculate a smooth, two-dimensional PES for CO on MgO, from which we extract the vibrational frequencies and frequency shift of CO stretching upon adsorption, finding quantitative agreement with infrared spectroscopy.
Using the CCSD(T) PES, we assess the accuracy of lower-level wavefunction theories and various DFT protocols that combine commonly used exchange-correlation functionals and dispersion corrections.
We find that none of the tested DFT protocols can achieve high accuracy in both the binding energy and the vibrational frequency.
By contrast, MP2 reproduces the entire CCSD(T) PES quite accurately, while CCSD performs slightly worse but still qualitatively correctly.

The ability to generate a smooth PES with high accuracy is a longstanding challenge in computational surface science for benchmarking and improving lower-level theories like DFT~\cite{Schimka10NM,Boereboom13JCP,Tchakoua19JPCC,Smeets19JPCA,Kroes21PCCP} and more recently, for training high-quality machine learning potentials that enable accurate \textit{ab initio} molecular dynamics study of surface reactions~\cite{Zhang18PRL,Wen23PNAS,Zeng23NC}.
For insulating and semiconducting surfaces with weak electron correlation, the smooth and accurate PESs attainable through our periodic local CCSD(T) method are thus very promising to fulfill both goals.
But for chemistry on the surface of metals or strongly correlated solids, for which CCSD(T) is inapplicable~\cite{Motta20PRX,Church21JCP,Lee22JCTC,Mihm21NCS,Neufeld23PRL,Masios23PRL}, new methods are sorely needed.

\section*{Conflict of interest}
The authors declare no competing conflicts of interest.

\section*{Acknowledgments}

This work was supported by the National Science Foundation under Grant No.~CHE-1848369 and by the Columbia Center for Computational Electrochemistry.
We acknowledge computing resources from Columbia University’s Shared Research Computing Facility project, which
is supported by NIH Research Facility Improvement Grant
1G20RR030893-01, and associated funds from the New York
State Empire State Development, Division of Science Technology and Innovation (NYSTAR) Contract C090171, both
awarded April 15, 2010.

\bibliography{refs}

\end{document}

\maketitle

    \vspace{3em}

    \noindent
    Note:
    \begin{enumerate}
        \item figures and equations appearing in the main text will be referred to as ``Fig.~Mxxx'' and ``Eq.~Mxxx'' in this Supplementary Material document.
        \item Further supplementary data can be found in the following Github repository
        \begin{center}
            \url{https://github.com/hongzhouye/supporting_data/tree/main/2023/arXiv%3A2309.14651}
        \end{center}
    \end{enumerate}

    \clearpage

    %\doublespacing
    \tableofcontents
    %\singlespacing

    \clearpage

    \section{Optimized geometries}

    The geometry optimization is performed at the DFT@PBE+D3~\cite{Perdew96PRL,Grimme10JCP} level of theory using Quantum Espresso~\cite{Giannozzi09JPCM,Giannozzi17JPCM} and the PAW pseudopotential~\cite{Blochl94PRB} (the $[2s^2 2p^6]$ of Mg is treated as valence electrons).
    The bulk MgO lattice is first optimized, from which the (001) surface is built and then optimized both with and without the CO adsorbate.
    A vacuum of $16$~\AA{} is employed.
    MgO slab models with two, three, and four atomic layers are considered in this work.
    During the optimization, atoms in the first layer of the two-layer slab model and the first two layers of both the three-layer and the four-layer models are allowed to relax, while atoms in the remaining layer(s) are kept fixed to their bulk optimized positions.
    The optimized structures are provided in the Github repository and summarized in \cref{tab:dft}.
    Our choice of DFT method (PBE+D3) is seen to well-reproduce the bulk MgO structure from both ref~\citenum{Shi23Chemrxiv} obtained computationally using DFT@revPBE+D4~\cite{Zhang98PRL,Caldeweyher17JCP} and ref~\citenum{Lazarov05PRB} obtained using XPD experiments.
    Due to the weak perturbation of CO to the surface, the surface structure is also well-predicted by PBE+D3, resulting in a geometry relaxation energy $\Delta_{\textrm{geom}}$ close to that in ref~\citenum{Shi23Chemrxiv} and several previous reports~\cite{Boese13PCCP,Maristella19JCTC}.
    However, PBE+D3 significantly underestimates the Mg-C distance and overestimates the adsorption energy.
    For this reason, in the main text we manually adjust the Mg-C distance to be $2.460$~\AA{} to match that from \cite{Shi23Chemrxiv} to facilitate the comparison.

    \begin{table}[!h]
        \centering
        \caption{Comparison of computational and experimental geometric and energetic parameters.}
        \label{tab:dft}
        % \begin{ruledtabular}
        \begin{tabular}{lccccl}
            \hline\hline
                & Lat.~const.~(\AA{})
                & $d(\textrm{Mg-C})$~(\AA{})
                & $\Delta_{\textrm{geom}}$~(kJ/mol)
                & $E_{\textrm{ads}}$~(kJ/mol)
                & Reference \\
            \hline
            PBE+D3
                & $4.222$ & $2.377$ & $1.1$ & $-29.5$ & this work  \\
            revPBE+D4
                & $4.220$ & $2.460$ & $0.8$ & $-20.0$ & ref~\citenum{Shi23Chemrxiv} \\
            Experiment
                & $4.217$ & & & & ref~\citenum{Lazarov05PRB}    \\
            \hline
        \end{tabular}
        % \end{ruledtabular}
    \end{table}

    \section{DFT single point calculations}

    All other DFT calculations, including the binding curves shown in FIG.~\fakeref{M2} and the vibrational frequencies shown in FIG.~\fakeref{M3} (which are in turn derived from the PESs shown in \cref{fig:TBF} below), were performed based on the optimized geometries obtained above.
    The non-dispersion part of the DFT energy was calculated using PySCF.
    The TS, XDM, and MBD dispersion corrections were calculated using Quantum Espresso with the norm-conserving Hartwigesen-Goedecker-Hutter pseudopotential~\cite{Goedecker96PRB,Hartwigsen98PRB}.
    The D2 and D3 dispersion corrections were calculated using the \texttt{simple-dftd3} code (\href{https://github.com/dftd3/simple-dftd3}{https://github.com/dftd3/simple-dftd3}).
    The D4 dispersion correction was calculated using the \texttt{dftd4} code (\href{https://github.com/dftd4/dftd4}{https://github.com/dftd4/dftd4}).

    % To allow calculating the dispersion corrections using the exchange-hole dipole moment~\cite{Becke05JCP} (XDM) method and the many-body dispersion~\cite{Tkatchenko12PRL} (MBD) method in Quantum Espresso, we use the norm-conserving Hartwigesen-Goedecker-Hutter (HGH) pseudopotential~\cite{Goedecker96PRB,Hartwigsen98PRB} for generating the binding curves in FIG.~\fakeref{M2}.
    % We verified that for the PBE functional, the calculated interaction energy from using the PAW and the HGH pseudopotentials differs by less than $10$~meV.
    % The XDM, PAW, and DFT-D2~\cite{Grimme06JCC} dispersion corrections were evaluated using Quantum Espresso.
    % The DFT-D3~\cite{Grimme10JCP} and DFT-D4~\cite{Caldeweyher17JCP,Caldeweyher19JCP,Caldeweyher20PCCP} dispersion corrections were evaluated using the open-source codes developed by the Grimme's group~\cite{DFTD3github,DFTD4github}

    \section{Wavefunction single point calculations}

    \subsection{Estimating errors from Gaussian basis sets and pseudopotentials}

    \begin{table}[!h]
        \centering
        \caption{Summary of the errors due to the employed Gaussian basis sets and pseudopotentials.}
        \label{tab:error_est}
        % \begin{ruledtabular}
        \begin{tabular}{ll}
            \hline\hline
            Source & Estimated error (kJ/mol)    \\
            \hline
            Diffuse functions & $0.1$  \\
            Pseudopotentials & $<0.1$ \\
            Semi-core electron correlation of Mg & $0.2$ \\
            \hline
            Total & $0.3$   \\
            \hline
        \end{tabular}
        % \end{ruledtabular}
    \end{table}

    In addition to the errors arising from TDL and CBS limit extrapolation as discussed in the Sec.~\fakeref{IIIA}, there are three major sources of errors in the correlated calculations of $E_{\textrm{int}}$:
    \begin{enumerate}
        \item Basis sets augmentation with diffuse functions
        \item Pseudopotentials
        \item Mg semicore electron correlation
    \end{enumerate}
    Our final estimates of these errors are summarized in \cref{tab:error_est} and detailed in what follows.
    All three sources combined give an error bar of about $0.3$~kJ/mol, which is added to the MP2 error bar discussed in the main text.

    \begin{figure}[!h]
        \centering
        \includegraphics[width=0.4\linewidth]{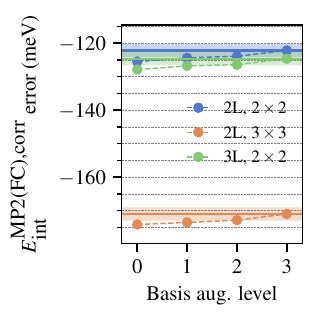}
        \caption{The correlation part of frozen-core MP2 interaction energy extrapolated to the CBS limit using TZ and QZ results for different choices of the basis set augmentation and slab models of different sizes.
        The shaded area indicates $\pm 0.2$~kJ/mol from the results obtained using level 3 basis set augmentation.}
        \label{fig:gtoaug_err}
    \end{figure}

    \noindent
    \textbf{Basis sets augmentation with diffuse functions}.
    For weak interactions, the regular Gaussian basis sets need to be augmented by diffuse functions for correlated calculations.
    We augment the GTH-cc-pV$X$Z basis sets~\cite{Ye22JCTC} (referred to as $X$Z henceforth) with one diffuse function per angular momentum channel to obtain the corresponding aug-GTH-cc-pV$X$Z basis sets (referred to as a$X$Z henceforth) for Mg, C, and O using the same protocol reported in ref~\citenum{Ye22JCTC}.
    We generated fitting basis sets for both the augmented and the non-augmented Gaussian basis sets used in this work.
    All basis set data can be found in the Github repository.

    Directly applying the a$X$Z basis sets to all atoms is unnecessary and also results in numerical problems due to the basis set linear dependency caused by the diffuse functions.
    We thus use a$X$Z basis sets only for the atoms near the adsorption center and monitor the convergence of the calculated interaction energy.
    Specifically, we define four levels of augmentation:
    \begin{enumerate}
        \item Level 0: using $X$Z basis sets for all atoms.
        \item Level 1: using a$X$Z basis sets for the CO molecule.
        \item Level 2: in addition to level 1, using a$X$Z basis sets for the surface Mg atom that binds CO.
        \item Level 3: in addition to level 2, using a$X$Z basis sets for the four surface O atoms next to the Mg atom that binds CO.
    \end{enumerate}
    In \cref{fig:gtoaug_err}, we examine the effect of different augmentation levels for the MP2 interaction energy extrapolated to the CBS limit using TZ and QZ results for slabs of different size.
    We use level 3 in our final calculations and take the difference between the results from level 2 and level 3 as our estimated error, which is about $0.2$~kJ/mol.

    \begin{table}[!h]
        \centering
        \caption{HF and MP2 interaction energy for $2\times2$ and $3\times3$ surfaces calculated using both the GTH pseudopotential (optimized for HF) and the all-electron potential.
        The $1s^2$ core electrons in all atoms (Mg, C, and O) are frozen in the all-electron MP2 calculations to match the number of electrons in the GTH pseudopotential.
        All calculations were performed using the level 3 basis set augmentation scheme.}
        \label{tab:pp_vs_alle}
        % \begin{ruledtabular}
        \begin{tabular}{llrrrr}
            \hline\hline
            Surface & Basis set & $E_{\textrm{int}}^{\textrm{HF,All-e}}$ & $E_{\textrm{int}}^{\textrm{HF,GTH}}$ & $E_{\textrm{int}}^{\textrm{MP2,All-e}}$ & $E_{\textrm{int}}^{\textrm{MP2,GTH}}$ \\
            \hline
            % \multirow{4}*{$2 \times 2$}
            %  & DZ & $50.5$ & $47.9$ & $-37.5$ & $-39.0$ \\
            %  & TZ & $40.0$ & $40.0$ & $-77.8$ & $-78.8$ \\
            %  & QZ & $38.1$ & $38.9$ & $-92.1$ & $-92.4$ \\
            %  \rowcolor{Gray}
            %  & CBS(TZ,QZ) & $38.1$ & $38.9$ & $-101.3$ & $-101.5$ \\
            % \hline
            % \multirow{4}*{$3 \times 3$}
            %  & DZ & $32.0$ & $28.6$ & $-100.5$ & $-103.4$ \\
            %  & TZ & $23.1$ & $23.0$ & $-144.0$ & $-145.1$ \\
            %  & QZ & $21.6$ & $22.5$ & $-159.1$ & $-159.5$ \\
            %  \rowcolor{Gray}
            %  & CBS(TZ,QZ) & $21.6$ & $22.5$ & $-169.0$ & $-169.6$ \\
            %  \hline
            \multirow{4}*{$2 \times 2$}
                & DZ & $4.9$ & $4.6$ & $-3.6$ & $-3.8$ \\
                & TZ & $3.9$ & $3.9$ & $-7.5$ & $-7.6$ \\
                & QZ & $3.7$ & $3.8$ & $-8.9$ & $-8.9$ \\
             \rowcolor{Gray}
                & CBS(TZ,QZ) & $3.7$ & $3.8$ & $-9.8$ & $-9.8$ \\
            \hline
            \multirow{4}*{$3 \times 3$}
                & DZ & $3.1$ & $2.8$ & $-9.7$ & $-10.0$ \\
                & TZ & $2.2$ & $2.2$ & $-13.9$ & $-14.0$ \\
                & QZ & $2.1$ & $2.2$ & $-15.4$ & $-15.4$ \\
             \rowcolor{Gray}
                & CBS(TZ,QZ) & $2.1$ & $2.2$ & $-16.3$ & $-16.4$ \\
             \hline
        \end{tabular}
        % \end{ruledtabular}
    \end{table}

    \noindent
    \textbf{Pseudopotential errors}.
    The error introduced by using the GTH pseudopotentials is examined by comparison with calculations using the all-electron potential and Dunning's (aug-)cc-pV$X$Z basis sets~\cite{Dunning89JCP,Kendall92JCP}.
    The results are presented in \cref{tab:pp_vs_alle} for different basis set size and surface size.
    We see that the at the CBS limit, the error caused by using the GTH pseudopotentials is less than $0.1$~kJ/mol for both HF and MP2.

    \begin{table}[!h]
        \centering
        \caption{HF and MP2 interaction energy for a $2\times2$ surfaces calculated using both the regular GTH-cc-pV$X$Z basis sets and the GTH-cc-pCV$X$Z basis sets with the GTH-HF pseudopotential.
        No electrons are frozen in the MP2 calculations.
        All calculations were performed using the level 3 basis set augmentation scheme.}
        \label{tab:core_err}
        % \begin{ruledtabular}
        \begin{tabular}{llrrrr}
            \hline\hline
            Surface & Basis set & $E_{\textrm{int}}^{\textrm{HF}}$ & $E_{\textrm{int}}^{\textrm{HF,core}}$ & $E_{\textrm{int}}^{\textrm{MP2}}$ & $E_{\textrm{int}}^{\textrm{MP2,core}}$ \\
            \hline
            % \multirow{4}*{$2 \times 2$}
            %  & DZ & $50.5$ & $50.1$ & $-37.5$ & $-39.3$ \\
            %  & TZ & $40.0$ & $39.4$ & $-77.8$ & $-80.6$ \\
            %  & QZ & $38.1$ & $37.7$ & $-92.1$ & $-94.3$ \\
            %  \rowcolor{Gray}
            %  & CBS(TZ,QZ) & $38.1$ & $37.7$ & $-101.3$ & $-103.1$ \\
            %  \hline
            \multirow{4}*{$2 \times 2$}
                & DZ & $4.9$ & $4.8$ & $-3.6$ & $-3.8$ \\
                & TZ & $3.9$ & $3.8$ & $-7.5$ & $-7.8$ \\
                & QZ & $3.7$ & $3.6$ & $-8.9$ & $-9.1$ \\
             \rowcolor{Gray}
                & CBS(TZ,QZ) & $3.7$ & $3.6$ & $-9.8$ & $-9.9$ \\
             \hline
        \end{tabular}
        % \end{ruledtabular}
    \end{table}

    \noindent
    \textbf{Correlation from semicore electrons of Mg}.
    The $[2s^2 2p^6]$ semicore electrons of Mg may contribute significantly to the interaction energy.
    To quantify this, we augment our (aug-)GTH-cc-pV$X$Z with core-correlating functions from the all-electron cc-pCV$X$Z basis sets~\cite{Woon95JCP}.
    The interaction energy evaluated with and without the core-correlating functions are listed in \cref{tab:core_err}.
    We see that at the CBS limit the error due to using the regular (aug-)GTH-cc-pV$X$Z basis sets is about $0.2$~kJ/mol for the MP2 interaction energy.

    \subsection{Convergence with the MgO slab thickness}

    \begin{table}[!h]
        \centering
        \caption{HF and MP2 interaction energy for a $2\times2$ surfaces with different slab thickness.
        No electrons are frozen in the MP2 calculations.
        All calculations were performed using the level 3 basis set augmentation scheme.}
        \label{tab:slab_err}
        % \begin{ruledtabular}
        \begin{tabular}{llrrrr}
            \hline\hline
            Surface & Basis set & $E_{\textrm{int}}^{\textrm{HF,2L}}$ & $E_{\textrm{int}}^{\textrm{HF,3L}}$ & $E_{\textrm{int}}^{\textrm{MP2,2L}}$ & $E_{\textrm{int}}^{\textrm{MP2,3L}}$ \\
            \hline
            % \multirow{4}*{$2 \times 2$}
            %  & DZ & $50.5$ & $50.8$ & $-37.5$ & $-42.3$ \\
            %  & TZ & $40.0$ & $38.8$ & $-77.8$ & $-82.1$ \\
            %  & QZ & $38.1$ & $38.0$ & $-92.1$ & $-95.2$ \\
            %  \rowcolor{Gray}
            %  & CBS(TZ,QZ) & $38.1$ & $38.0$ & $-101.3$ & $-104.2$ \\
            %  \hline
            \multirow{4}*{$2 \times 2$}
                & DZ & $4.9$ & $4.9$ & $-3.6$ & $-4.1$ \\
                & TZ & $3.9$ & $3.7$ & $-7.5$ & $-7.9$ \\
                & QZ & $3.7$ & $3.7$ & $-8.9$ & $-9.2$ \\
             \rowcolor{Gray}
                & CBS(TZ,QZ) & $3.7$ & $3.7$ & $-9.8$ & $-10.1$ \\
             \hline
        \end{tabular}
        % \end{ruledtabular}
    \end{table}

    In \cref{tab:slab_err}, we compare the HF and MP2 interaction energy calculated for the 2-layer and 3-layer slab models.
    At the CBS limit, the difference between using the two slab models is negligible for HF and about $0.3$~kJ/mol for MP2.
    Given the small energy differences and the higher computational expenses of using the thicker slabs, we use the 2-layer slab model for all calculations reported in the main text.

    \subsection{Transferability of the $\Delta^{\textrm{CC(FC)}}(\eta)$ term}

    \begin{figure}[!h]
        \centering
        \includegraphics[width=0.4\linewidth]{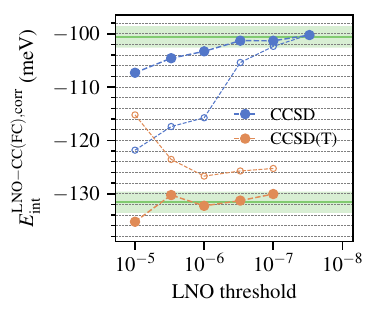}
        \caption{Same plot as FIG.~\fakeref{M1E} but for a $3\times3$ surface with the DZ basis set, with hollow and filled circles denoting uncorrected and corrected results, respectively.}
        \label{fig:lno_cano_331_dz}
    \end{figure}

    To examine the transferability of the correction term, $\Delta^{\textrm{CC(FC)}}(\eta)$, defined in eqn~(\fakeref{M7}) economically evaluated using the DF basis set and $2\times2$ surface, in \cref{fig:lno_cano_331_dz} we show the $\eta$-convergence of the LNO-CCSD/CCSD(T) interaction energy for a larger surface of size $3 \times 3$.
    We see that $\Delta^{\textrm{CC(FC)}}(\eta)$ is as effective at accelerating the $\eta$-convergence for the larger surface as for the smaller one shown in FIG.~\fakeref{M1E} for both LNO-CCSD and LNO-CCSD(T).
    We hence apply $\Delta^{\textrm{CC(FC)}}(\eta)$ to all the LNO-CCSD/CCSD(T) calculations reported in the main text.

    \section{Potential energy surface and vibrational frequency}

    \subsection{PES fitting}

\begin{figure}[!htbp]
        \centering
        \includegraphics[width=0.95\linewidth]{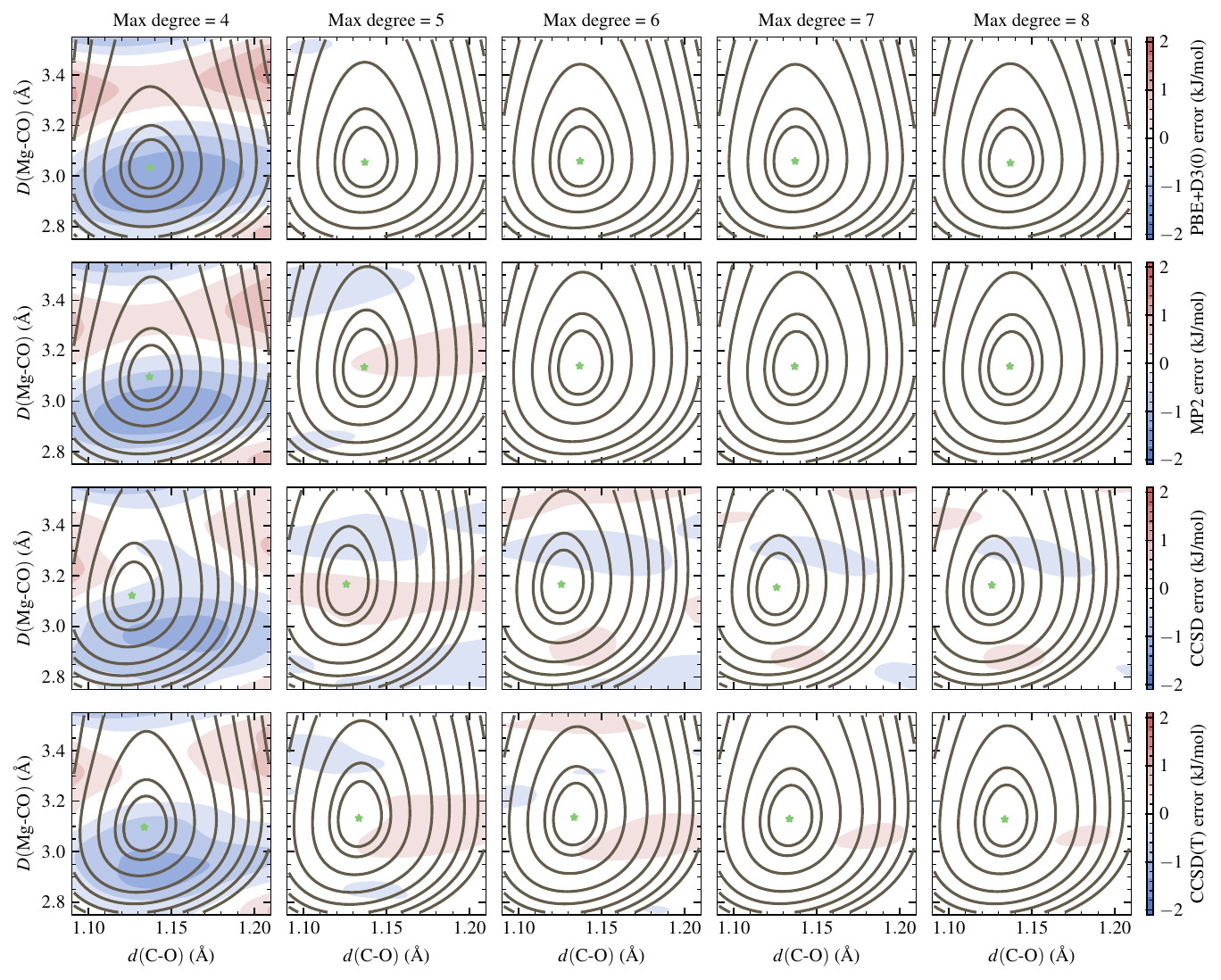}
        \caption{Error of the polynomial fitted PES (\ref{eq:twodim_PES}) of different polynomial order for PBE+D3(0), MP2, CCSD, and CCSD(T).}
        \label{fig:pes_conv}
    \end{figure}

\begin{figure}[!htbp]
        \centering
        \includegraphics[width=0.5\linewidth]{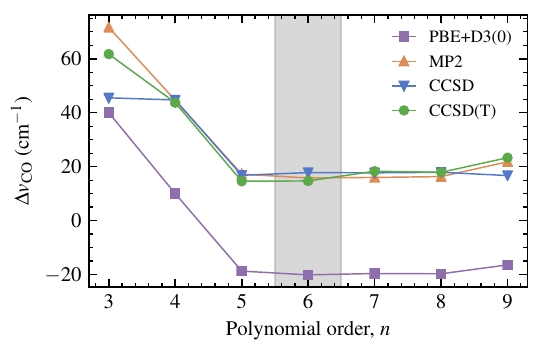}
        \caption{Convergence of the vibrational frequency shift with the polynomial order $n$ used to fit the two-dimensional PES for evaluating $\nu_{\textrm{CO(ads)}}$.}
        \label{fig:freq_conv}
    \end{figure}

    For each method discussed in Sec.~\fakeref{IIIC}, we perform $130$ single point energy calculations for $d(\textrm{C-O})$ from $0.982$ to $1.432$~\AA{} with an increment of $0.05$~\AA{} and $D(\textrm{Mg-CO})$ from $2.630$ to $3.830$~\AA{} with an increment of $0.1$~\AA{}.
    We fit the energy data using an $n$-th order polynomial
    \begin{equation}    \label{eq:twodim_PES}
        E(d,D;n)
            = \sum_{m = 0}^{n} \sum_{p=0}^{m} a^{(n)}_{mp} d^p D^{m-p}
    \end{equation}
    From the fitted PES, the vibrational frequency can be calculated by diagonalizing the analytical Hessian evaluated at equilibrium $(d_0, D_0)$.
    In \cref{fig:pes_conv}, we show the fitting error of polynomial order $n = 4$ -- $8$ for the PBE+D3(0), MP2, CCSD, and CCSD(T) PES, calculated using the QZ basis set and the $3 \times 3$ surface.
    The vibrational frequency shift calculated for these PESs are plotted in \cref{fig:freq_conv}.
    Based on these results, we identify $n = 6$ as the optimal choice that achieves both accurate fitting (with a maximum fitting error less than $0.5$~kJ/mol) and stable frequency shift.
    We separately verified that fitting the one-dimensional PES for the free CO molecule with a $6$-th order polynomial leads to $\nu_{\textrm{CO(g)}}$ that agrees with that calculated using ORCA~\cite{Neese20JCP} based on analytical Hessians.
    In the following we fix $n = 6$ and study the convergence with respect to basis set size and surface size.

    \subsection{Basis set, surface size, and LNO subspace}

    \Cref{tab:freq_basis_surf_conv} shows the convergence of PBE and MP2 $\Delta \nu_{\textrm{CO}}$ with respect to the surface size and basis set size.
    From the table, we determine that the QZ basis set and $3 \times 3$ surface is suitable for all calculations.

    \begin{table}[!h]
        \centering
        \caption{CO stretching frequency shift (in $\wavenumber$) upon adsorption calculated using PBE and MP2 for different surface size and basis sets.
        QZ basis set and $3 \times 3$ surface are sufficient for a converged frequency calculation.}
        \label{tab:freq_basis_surf_conv}
        \begin{tabular}{llll}
            \hline\hline
            Surface & Basis set & PBE & MP2 \\
            \hline
            $3 \times 3$ & DZ & $-12.7$ & $24.9$  \\
            $3 \times 3$ & TZ & $-20.6$ & $17.2$   \\
            \rowcolor{Gray}
            $3 \times 3$ & QZ & $-20.2$ & $15.8$   \\
            $4 \times 4$ & DZ & $-12.7$ & $24.9$   \\
            \hline
        \end{tabular}
    \end{table}

    \Cref{tab:freq_lno_conv} shows the $\eta$-convergence of the LNO-CCSD and LNO-CCSD(T) predicted frequency shift, from which we see that $\eta = 3 \times 10^{-6}$ converges $\Delta \nu_{\textrm{CO}}$ to within one wavenumber.
    The final CCSD/CCSD(T) frequency shift reported in the main text is calculated using $3\times3$/QZ energy obtained by the $3\times3$/TZ energy composite corrected by the difference between $2\times2$/TZ and $2\times2$/QZ.

    \begin{table}[!h]
        \centering
        \caption{$\eta$-convergence of CO stretching frequency shift (in $\wavenumber$) upon adsorption calculated using LNO-CCSD and LNO-CCSD(T) for two choices of surface size/basis set size: $2 \times 2$/TZ and $3 \times 3$/DZ.
        $\eta = 3 \times 10^{-6}$ is seen to reach convergence within one wavenumber for both cases.}
        \label{tab:freq_lno_conv}
        \begin{tabular}{llll}
            \hline\hline
            Surface/basis & $\eta$ & CCSD & CCSD(T) \\
            \hline
            \multirow{4}*{$2 \times 2$/TZ}
                & 1E-5 & $39.6$ & $19.7$  \\
                & 3E-6 & $29.1$ & $23.5$  \\
                & 1E-6 & $28.4$ & $24.4$  \\
                & 3E-7 & $28.5$ & $22.9$  \\
            \hline
            \multirow{3}*{$3 \times 3$/DZ}
                & 1E-5 & $27.6$ & $23.4$  \\
                & 3E-6 & $26.0$ & $23.2$  \\
                & 1E-6 & $25.1$ & $23.7$  \\
            \hline
        \end{tabular}
    \end{table}

    \subsection{Final PES}

\begin{figure}[!h]
        \centering
        \includegraphics[width=0.8\linewidth]{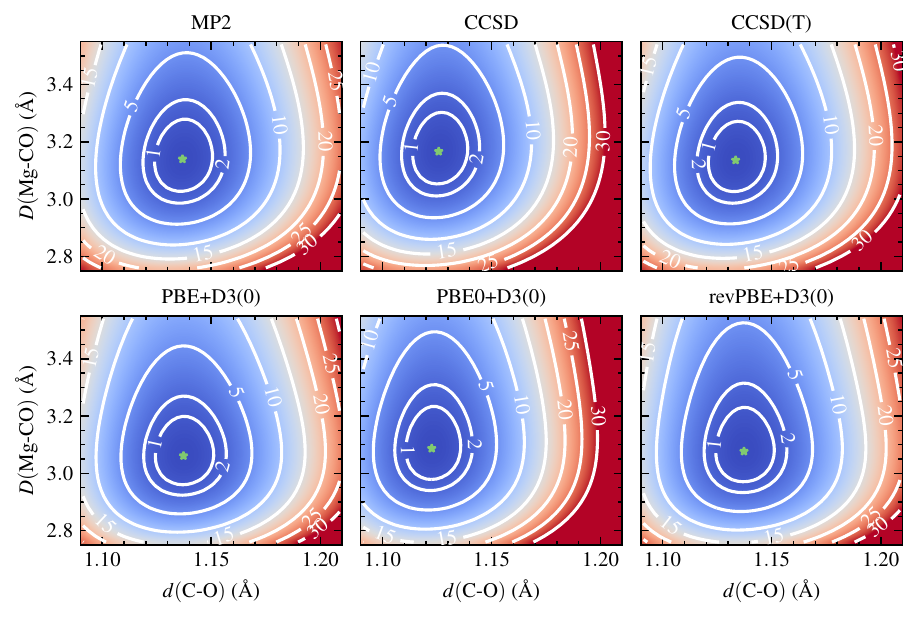}
        \caption{Same PES contour plot as in FIG.~\fakeref{M3A} for different methods.}
        \label{fig:pes_final}
    \end{figure}

    The final PES obtained using the protocols discussed above are shown in \cref{fig:pes_final} for different methods.

    \clearpage

    \bibliography{refs_si}